\documentclass[preprint,12pt]{elsarticle}

\usepackage{amssymb,amsmath}

\journal{arXiv}

\begin{document}

\begin{frontmatter}



\title{A gamma variate generator with shape parameter less than unity}


\author{Seiji Zenitani}

\affiliation{organization={Space Research Institute, Austrian Academy of Sciences},
            addressline={Schmiedlstra{\ss}e 6}, 
            city={Graz},
            postcode={8042}, 
            country={Austria}}

\begin{abstract}
Algorithms for generating random numbers that follow a gamma distribution
with shape parameter less than unity are proposed. 
Acceptance-rejection algorithms are developed, 
based on the generalized exponential distribution.
The squeeze technique is applied to our method, and then
piecewise envelope functions are further considered. 
The proposed methods are excellent in acceptance efficiency 
and promising in speed.
\end{abstract}




\end{frontmatter}

\section{Introduction}

The gamma distribution is one of
the most important probability distributions in statistics.
It has applications in a broad range of fields
in natural sciences, engineering, and social sciences. 
The probability density is defined by 
\begin{align}
f_{\rm GA}(x;\alpha,\lambda)
=
\frac{1}{\lambda^\alpha \Gamma(\alpha)} x^{\alpha-1}e^{-x/\lambda}
~~
(x \ge 0)
\label{eq:gamma}
\end{align}
where $\alpha$ ($>0$) is the shape parameter,
$\lambda$ ($>0$) the scale parameter, and
$\Gamma(x)$ the gamma function.

Generating a gamma variate,
a random number drawn from a gamma distribution,
is an important and fundamental problem in statistical computing.
Owing to its importance,
various numerical algorithms have been developed
since the early days of computer age
\citep{AD74,best83,devroye86,mt00,kundu07,tanizaki08,yotsuji10,kroese11}.

The nature of the gamma distribution is controlled
by the shape parameter $\alpha$.
Accordingly, we need to use different gamma variate generators.
When $1 < \alpha $, the distribution function increases from zero at $x=0$,
has a maximum at $x=(\alpha-1)\lambda$, and then decays as $x$ further increases.
To generate a gamma variate with $\alpha > 1$,
probably \citet{mt00}'s algorithm is one of the best,
as detailed in literature \citep{mt00,yotsuji10,kroese11}.
When $\alpha=1$, the distribution is reduced to the exponential distribution, and it is easy to generate a variate.

When $0<\alpha<1$, which is the range of our interest,
the gamma distribution monotonically decreases
from positive infinity at $x=0$ to zero at $x \rightarrow \infty$.
To generate a gamma variate for $0<\alpha<1$,
various acceptance-rejection methods have been developed
\citep{AD74,best83,devroye86,kundu07}.
Combining power-law and exponential distributions,
\citet{AD74} proposed a piecewise rejection method.
\citet{best83} extended the \citet{AD74} method,
by adjusting a switching point and introducing a squeeze technique.
\citet{devroye86} developed a different rejection method,
based on the exponential power distribution.
As of 2024, it is employed by a Python package, {\tt NumPy} \citep{scipy}.
Using a generalized exponential distribution \citep{gupta99},
\citet{kundu07} developed a rejection method and its piecewise extensions.
Finally, \citet{tanizaki08} developed a ratio-of-uniforms method,
which works either for $0<\alpha<1$ and for $1<\alpha$.

In this note, we propose efficient gamma generators for $0<\alpha<1$,
advancing an earlier study \citep{kundu07}. 
In Section \ref{sec:base}, 
we develop an acceptance-rejection algorithm,
using the generalized exponential distribution. 
In Section \ref{sec:squeeze}, 
we construct fractional functions for the squeeze technique. 
In Section \ref{sec:piecewise},
we discuss piecewise envelope functions. 
In Section \ref{sec:test}, we evaluate
the performance of the proposed and previous methods.
Section \ref{sec:discussion} contains discussions and the summary.

\section{Base algorithm}
\label{sec:base}

We construct a base algorithm to generate
a random variate drawn from the gamma distribution with shape $0<\alpha<1$.
We focus on the $\lambda=1$ case,
because we can obtain the results for $\lambda \ne 1$,
by multiplying the outputs by $\lambda$.
\begin{align}
f_{\rm GA}(x;\alpha,1) &= \frac{1}{\Gamma(\alpha)} x^{\alpha-1}e^{-x}
\label{eq:GA}
\end{align}
In line of \citet{kundu07},
we employ the generalized exponential (GE) distribution \citep{gupta99},
\begin{align}
f_{\rm GE}(x; \alpha, \lambda) = \dfrac{\alpha}{\lambda} (1-e^{-x/\lambda})^{\alpha-1} e^{- x/\lambda}
~~~~(x \ge 0)
\label{eq:GE} 
\end{align}
where $\alpha$ ($>0$) and $\lambda$ ($>0$) are the shape and scale parameters.
Note that Refs.~\cite{gupta99} and \cite{kundu07} use different conventions for the scale parameter, and we employ \citet{gupta99}'s.
The cumulative distribution function (CDF) of the GE distribution is
\begin{align}
F_{\rm GE}(x; \alpha, \lambda) = (1-e^{-x/\lambda})^{\alpha}
\label{eq:GE_CDF} 
\end{align}
Hereafter we consider the GE distribution with $\lambda=1$.
We generate a uniform random variate $U_1$ in the $(0,1)$ range:
\begin{align}
U_1 \sim U(0,1)
\end{align}
Equating $U_1$ with \eqref{eq:GE_CDF} with $\lambda=1$,
we find that
a random variate $x$ that follows the generalized exponential distribution
$f_{\rm GE(\alpha,1)}$ 
is drawn by
\begin{align}
x \leftarrow -\log( 1 - U_1^{1/\alpha} )
\label{eq:XGE}
\end{align}

To help our discussion, we use a parameter $\beta$,
a function $g(x)$, and a rejection function $R_1(x)$.
\begin{align}
\beta &\equiv 1-\alpha,~~~
g(x) \equiv \dfrac{1-e^{-x}}{x} \label{eq:g} \\
R_1(x) &\equiv \left( \dfrac{x}{1-e^{-x}} \right)^{\alpha-1} = \left( g(x) \right)^{\beta}
\label{eq:R1}
\end{align}
The function $g(x)$ is continuous at $x=0$.
When $x \ge 0$, as $g(0)=1$ and $g'(x)=(1+x-e^x)/(x^2e^x) \le 0$, we find $0 < g(x) \le 1$.
Since $0<\beta<1$, we find
\begin{align}
0< R_1(x) \le  1
\label{eq:rej}
.
\end{align}
All the equal signs are met when $x=0$.

Comparing the GE \eqref{eq:GE} and gamma distributions \eqref{eq:GA},
we rewrite \eqref{eq:GA} as follows.
\begin{align}
f_{\rm GA}(x;\alpha,1)
= \frac{1}{\Gamma(\alpha+1)} R_1(x)~f_{\rm GE}(x;\alpha,1)
\label{eq:GA3}
\end{align}
Here, the relation $\alpha\Gamma(\alpha)=\Gamma(\alpha+1)$ is used. 
Eq.~\eqref{eq:GA3} suggests that
we can apply a rejection method to the GE-distributed variate $x$, 
to obtain a gamma-distributed variate.
Using another uniform variate $U_2 \sim U(0,1)$,
we evaluate the acceptance condition:
\begin{align}
U_2 \le R_1(x)
\label{eq:rej1}
\end{align}
or an equivalent inequality:
\begin{align}
(U_2)^{1/(1-\alpha)} x \le U_1^{1/\alpha}
\label{eq:rej2}
\end{align}
If either \eqref{eq:rej1} or \eqref{eq:rej2} is satisfied,
then we take this number $x$.
If it is not satisfied, we discard the number, and then
we go back to the beginning.
The final output follows a gamma distribution with shape $\alpha$.
These procedures are summarized in {Algorithm 1} in Table \ref{table:GE1}.
We need the total density $1/\Gamma(\alpha+1)$ to generate one gamma distribution. 

\begin{table}[tbhp]
\centering
\caption{A base algorithm of our generators for $0<\alpha<1$.
\label{table:GE1}}
\begin{tabular}{l}
\hline
{\bf Algorithm 1}\\
\hline
{\bf repeat} \\
~~~~generate $U_1, U_2 \sim U(0, 1)$ \\
~~~~$b \leftarrow  U_1^{1/\alpha}$,~~$x \leftarrow -\log(1 - b)$ \\
~~~~{\bf if}~ $U_2^{1/(1-\alpha)} x \le b$ ~~{\bf return} $x$ \\
{\bf end repeat} \\
\hline
\end{tabular}
\end{table}

\section{Squeeze technique}
\label{sec:squeeze}

We improve the speed of the algorithm by the so-called squeeze method.
We avoid the power and exponential operations
in the acceptance condition \eqref{eq:rej1},
by using the following inequalities for $x \ge 0$.
\begin{align}
\dfrac{4-\beta x}{4+\beta x}
\le R_1(x)
\le \dfrac{4+(1-\beta) x}{4+(1+\beta) x}
\label{eq:sandwitch1}
\end{align}

To show this, we start from several inequalities:
\begin{align}
e^{-x} & \ge \dfrac{2-x}{2+x}   ~~~~~ {\bf if} ~ x \ge 0
\label{eq:ineq1}
\\
x^{\beta} &\le \dfrac{(1-\beta)+(1+\beta)x}{(1+\beta)+(1-\beta)x} ~~~~~~ {\bf if} ~ 0 < x \le 1 {\rm~and~} 0 \le \beta \le 1
\label{eq:ineq3}
\\
x^{\beta} &\ge \dfrac{(1-\beta)+(1+\beta)x}{(1+\beta)+(1-\beta)x} ~~~~~~ {\bf if} ~ x \ge 1 {\rm~and~} 0 \le \beta \le 1
\label{eq:ineq4}
\\
x^\beta &\le 1+ \beta(x-1)  ~~~~~~ {\bf if} ~ 0 < x  {\rm~and~} 0 \le \beta \le 1
\label{eq:ineq5}
\end{align}
We can prove them in the following way.
\begin{align}
p(x) &\equiv \dfrac{2-x}{2+x}e^{x}
{\rm ~~which~tells~}
p(0)=1,~
p'(x) = \dfrac{-x^2 e^x}{(2+x)^2}<0
\\
q(x) &\equiv
\dfrac{(1-\beta)+(1+\beta)x}{(1+\beta)+(1-\beta)x}~ x^{-\beta} 
{\rm ~~which~tells~}
q(1) = 1,~
\nonumber \\
&~~~~~
q'(x) =
\dfrac{\beta (\beta^2-1) (1-x)^2 x^{-\beta-1}}{[(1+\beta)+(1-\beta)x ]^2}
\le 0
\\
r(x) &\equiv -x^\beta + 1+ \beta(x-1)
{\rm ~~which~tells~}
r(1) = 0,~
\nonumber \\
&~~~~~
r'(x) = \beta(1- x^{\beta-1}) < 0 {\rm~for~}x<1,~
r'(x) > 0 {\rm~for~}x>1
\end{align}

For the left inequality of \eqref{eq:sandwitch1},
we differentiate the rejection function
\begin{align}
R'_1(x)
&=
\beta \left( \frac{x-e^{x}+1}{x(e^{x}-1)} \right) R_1(x)
\end{align}
With help from the Taylor series $e^x = \sum_{n=0}^\infty \dfrac{x^n}{n!}$,
we find
\begin{align}
\frac{R'_1(x)}{R_1(x)}
= -\beta \dfrac{\sum_{n=2}^\infty \dfrac{x^n}{n!}}{ x\sum_{n=1}^\infty \dfrac{x^n}{n!}}
= -\beta \dfrac{\sum_{n=2}^\infty \dfrac{1}{n}\dfrac{x^n}{(n-1)!}}{ \sum_{n=2}^\infty \dfrac{x^n}{(n-1)!}}
\end{align}
In the right-hand side, all the terms in the numerator and in the denominator are positive,
but the numerator terms contain additional factor of $1/2, 1/3, \cdots$.
This indicates that the ratio of the numerator to the denominator is less than $1/2$.
Together with the $x=0$ case, we find
\begin{align}
\frac{R'_1(x)}{R_1(x)}
\ge -\frac{\beta}{2}
\end{align}
This suggests, together with \eqref{eq:ineq1},
\begin{align}
R_1(x)
\ge \exp\left(-\dfrac{\beta}{2}x\right)
\ge \dfrac{4-\beta x}{4+\beta x}
.
\label{eq:sq2}
\end{align}
For the right inequality of \eqref{eq:sandwitch1},
with help from \eqref{eq:R1}, \eqref{eq:ineq1} and \eqref{eq:ineq3}, we obtain
\begin{align}
R_1(x)
\le \left( \dfrac{2}{2+x} \right)^{\beta}
\le \dfrac{4+(1-\beta) x}{4+(1+\beta) x}
\label{eq:sq1}
\end{align}
Thus \eqref{eq:sq2} and \eqref{eq:sq1} prove \eqref{eq:sandwitch1}.

In practice, we utilize the inequalities \eqref{eq:sandwitch1} or equivalent logical expressions, before evaluating $R_1(x)$. 
Translating $\beta \rightarrow (1-\alpha)$ again,
we show the final algorithm in {Algorithm 2} (Table \ref{table:GE2}).

\begin{table}[tbhp]
\centering
\caption{A gamma generator with the squeeze technique.
\label{table:GE2}}
\begin{tabular}{l}
\hline
{\bf Algorithm 2}\\
\hline
{\bf repeat} \\
~~~~generate $U_1, U_2 \sim U(0, 1)$ \\
~~~~$b \leftarrow  U_1^{1/\alpha}$,~~$x \leftarrow -\log(1 - b)$ \\
~~~~{\bf if}~ $U_2 (4+(1-\alpha) x) \le (4+(\alpha-1)x)$ ~~{\bf return} $x$ \\
~~~~{\bf if}~ $U_2 (4+(2-\alpha) x) \le (4+\alpha x)$ {\bf then} \\
~~~~~~~~{\bf if}~ $U_2^{1/(1-\alpha)} x \le b$ ~~{\bf return} $x$ \\
{\bf end repeat} \\
\hline
\end{tabular}
\end{table}

\section{Piecewise envelope functions}
\label{sec:piecewise}

Next, we split the envelope function for the rejection method
into two parts.
Across the switching point at $x=s$ ($>0$),
we consider the GE distribution in the left
and an exponential tail in the right. 
With help from the CDF of the GE distribution (Eq.~\eqref{eq:GE_CDF}),
we rewrite the gamma distribution (Eq.~\eqref{eq:GA}) as follows.
\begin{align}
f_{\rm GA}(x; \alpha,1)
&=
\left\{
\begin{array}{ll}
S_L \cdot
R_1(x)
~
\dfrac{ f_{\rm GE}(x;\alpha,1) }{ F_{\rm GE}(s;\alpha,1) }
~~
& {\rm (for~ 0 \le x \le s)}
\\[10pt]
S_R \cdot
R_2(x)
~e^{-(x-s)}
& {\rm (for~x > s)}
\end{array}
\right.
\label{eq:split2}
\end{align}
Here, $S_L$ and $S_R$ are the densities of the left and right parts,
\begin{align}
S_L
\equiv
\dfrac{(1-e^{-s})^\alpha}{\Gamma(\alpha+1)}
,~~
S_R
\equiv
\dfrac{\alpha e^{-s} s^{\alpha-1}}{\Gamma(\alpha+1)}
,
\end{align}
and $R_2(x)$ is the second rejection function
\begin{align}
R_2(x) \equiv \left( \frac{x}{s} \right)^{\alpha-1}
\label{eq:R2}
\end{align}
which satisfies $0 < R_2(x) < 1$ for $x > s$. 
Equation \eqref{eq:split2} tells us that
we can apply the rejection method to the GE distribution with $R_1(x)$ in the left part, and that
we can apply the rejection method to the exponential density
with $R_2(x)$ in the right part.

Using a uniform variate $U_1$, when $U_1 \le S_L/(S_L+S_R)$,
we proceed to the GE part.
We generate the GE-distributed number in the range $[0,s]$.
Using a uniform variate $U_2$, the GE variate $x$ can be drawn by
\begin{align}
x \leftarrow
-\log \left( 1 - [F_{\rm GE}(s; \alpha,1)~U_2]^{1/\alpha} \right)
\end{align}
Then we evaluate the acceptance condition,
as discussed in Section \ref{sec:squeeze}.

In the right part,
we obtain an exponential variate from $U_2$,
\begin{align}
x \leftarrow s - \log U_2
\end{align}
and then we apply the rejection method.
We can similarly apply the squeeze technique to \eqref{eq:R2}.
For $x \ge 1$, from \eqref{eq:ineq4} and \eqref{eq:ineq5}, 
we obtain a squeeze relation
\begin{align}
\dfrac{1}{1+ \beta(x-1)}
\le
x^{-\beta}
\le 
\dfrac{(1+\beta)+(1-\beta)x}{(1-\beta)+(1+\beta)x}
\label{eq:sandwitch2}
\end{align}

Finally, we choose $s$. 
First, we look for $s$ that minimizes the total density
\begin{align}
S(\alpha, s) = {S_L+S_R}
=
\dfrac{(1-e^{-s})^{\alpha} + \alpha{s^{\alpha-1}}e^{-s}}{\Gamma(\alpha+1)}
.
\label{eq:den_s}
\end{align}
The solution can be found by solving
$
(1-e^{-s})^{\alpha-1} + (\alpha-1){s^{\alpha-2}} - {s^{\alpha-1}}
= 0
$
by a root finder.
In this case, we find an approximation by trial and error,
\begin{align}
s^{*} = 1.28 + 0.23\alpha
\label{eq:approx}
\end{align}
This gives a near-minimum $S(\alpha, s^{*})$
within an error of $< 1.5 \times 10^{-6}$. 
Another choice is $s=1$, in analogy with \citet{AD74}.
This simplifies several parameters in the algorithm.

These procedures are shown in {Algorithm 3} (Table \ref{table:GE3}).
The first three lines initialize coefficients, and then
the lines inside the loop generate the variate.
We reuse $(S_L+S_R)/S_L \cdot U_1$ or $(S_L+S_R)/S_R \cdot (U_1-S_L/(S_L+S_R))$ as a uniform variate, in order to reduce the total number of random variates. 
When $s=1$, we can further simplify the code, which is a trivial task.

\begin{table}[tbhp]
\centering
\caption{A method with the piecewise envelope functions.
\label{table:GE3}}
\begin{tabular}{l}
\hline
{\bf Algorithm 3}\\ 
\hline
$s \leftarrow 1.28 + 0.23 \alpha$,~~$t \leftarrow \exp(-s)$~~~~// $s=1,~t=1/e$\\
$S_L \leftarrow (1-t)^{\alpha}$,~~$S_R \leftarrow \alpha ~t~ s^{\alpha-1}$,~~$S \leftarrow S_L + S_R$ \\
$p_1 \leftarrow S_L / S$,~~$d_2 \leftarrow S / S_R$ \\
{\bf repeat} \\
~~~~generate $U_1, U_2 \sim U(0, 1)$ \\
~~~~{\bf if}~ $U_1 \le p_1$ {\bf then} \\
~~~~~~~~$b \leftarrow  (S U_1)^{1/\alpha}$,~~$x \leftarrow -\log(1 - b)$ \\
~~~~~~~~{\bf if}~ $U_2 (4+(1-\alpha) x) \le (4+(\alpha-1)x)$ ~~{\bf return} $x$ \\
~~~~~~~~{\bf if}~ $U_2 (4+(2-\alpha) x) \le (4+\alpha x)$ {\bf then} \\
~~~~~~~~~~~~{\bf if}~ $U_2^{1/(1-\alpha)} x \le b$ ~~{\bf return} $x$ \\
~~~~{\bf else} \\
~~~~~~~~$x \leftarrow s -\log d_2(U_1 - p_1)$,~~$y \leftarrow x/s$ \\
~~~~~~~~{\bf if}~ $U_2 (\alpha+(\alpha-1) y) \le 1$ ~~{\bf return} $x$ \\
~~~~~~~~{\bf if}~ $U_2 (\alpha+(2-\alpha) y) \le (2-\alpha + \alpha y)$ {\bf then} \\
~~~~~~~~~~~~{\bf if}~ $U_2 \le y^{\alpha-1}$ ~~{\bf return} $x$ \\
~~~~{\bf endif} \\
{\bf end repeat} \\
\hline
\end{tabular}
\end{table}


\section{Numerical tests}
\label{sec:test}

We have carried out several benchmark tests.
We wrote C codes for the eight methods:
1) the \citet{AD74} method, 
2) the \citet{best83} method,
3) the \citet{devroye86} method,
4) the \citet{kundu07} method (Algorithm 3 in their paper),
5) Algorithm 1 (Table \ref{table:GE1}),
6) Algorithm 2 (Table \ref{table:GE2}),
7) Algorithm 3 (Table \ref{table:GE3}) with $s=s^{*}$, and
8) Algorithm 3 with $s=1$.
In the last case, we have simplified the algorithm,
as mentioned in Sec. \ref{sec:piecewise}.
We use a uniform random generator ({\tt gsl\_rng\_uniform}) in the GNU Scientific Library.
We use Intel oneAPI Compiler ({\tt icx}) v2024.1 with the {\tt -lgsl -O0 -lm} option and the clang compiler ({\tt clang}) v14.0 with the {\tt -lgsl -lm} option on AMD Ryzen 5955 processor on Ubuntu Linux 24.04. 



\begin{figure}[htbp]
\centering
\includegraphics[width={\columnwidth}]{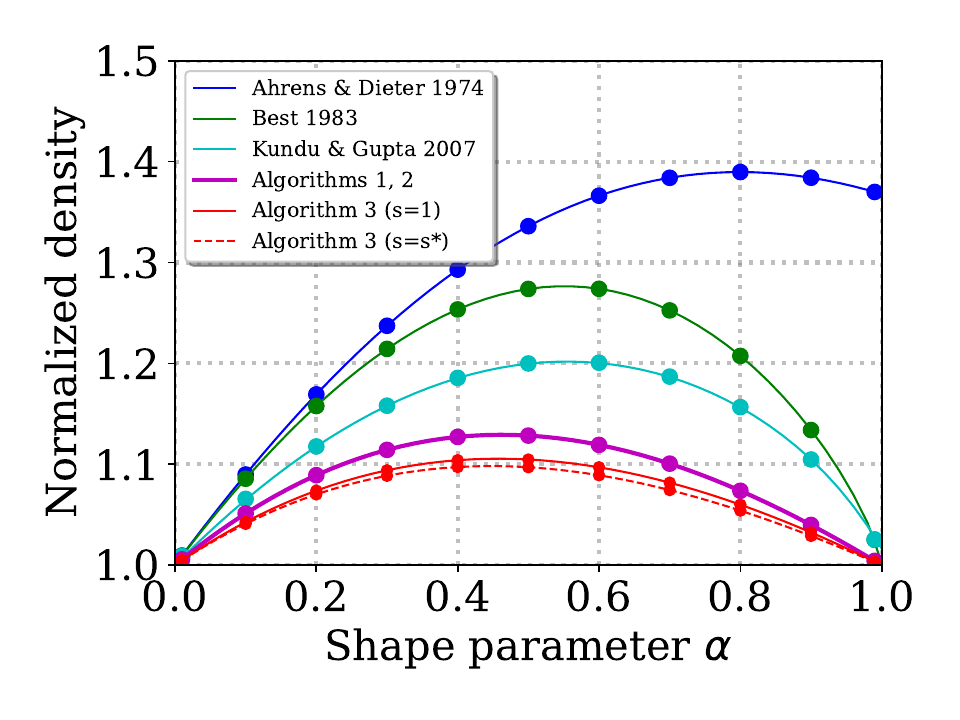}
\caption{
Normalized number densities to generate a gamma distribution.
Theoretical predictions (curves) and numerical results (circles) are shown,
as a function of $\alpha$.
\label{fig:density}}
\end{figure}

Figure \ref{fig:density} shows
the normalized number of random variates
to generate a gamma distribution,
as a function of $\alpha$.
The curves indicate theoretical predictions.
Algorithms 1 \& 2 require $1/\Gamma(\alpha+1)$,
the same number as the most efficient method to date,
the \citet{devroye86} method (not shown).
Algorithm 3's number is given by \eqref{eq:den_s}. 
The circles indicate our Monte Carlo results
for eleven shape parameters, $\alpha=0.01, 0.1, 0.2, 0.3, \cdots 0.9, 0.99$.
In each case, $10^8$ random numbers are generated.
They are in excellent agreement with the theories.
The proposed methods, Algorithm 3 in particular,
need fewer random numbers than most of the other algorithms.
The $s=s^{*}$ case is better than the $s=1$ case by $\lesssim 0.8$\%.

\begin{figure*}[htbp]
\centering
\includegraphics[width={\textwidth}]{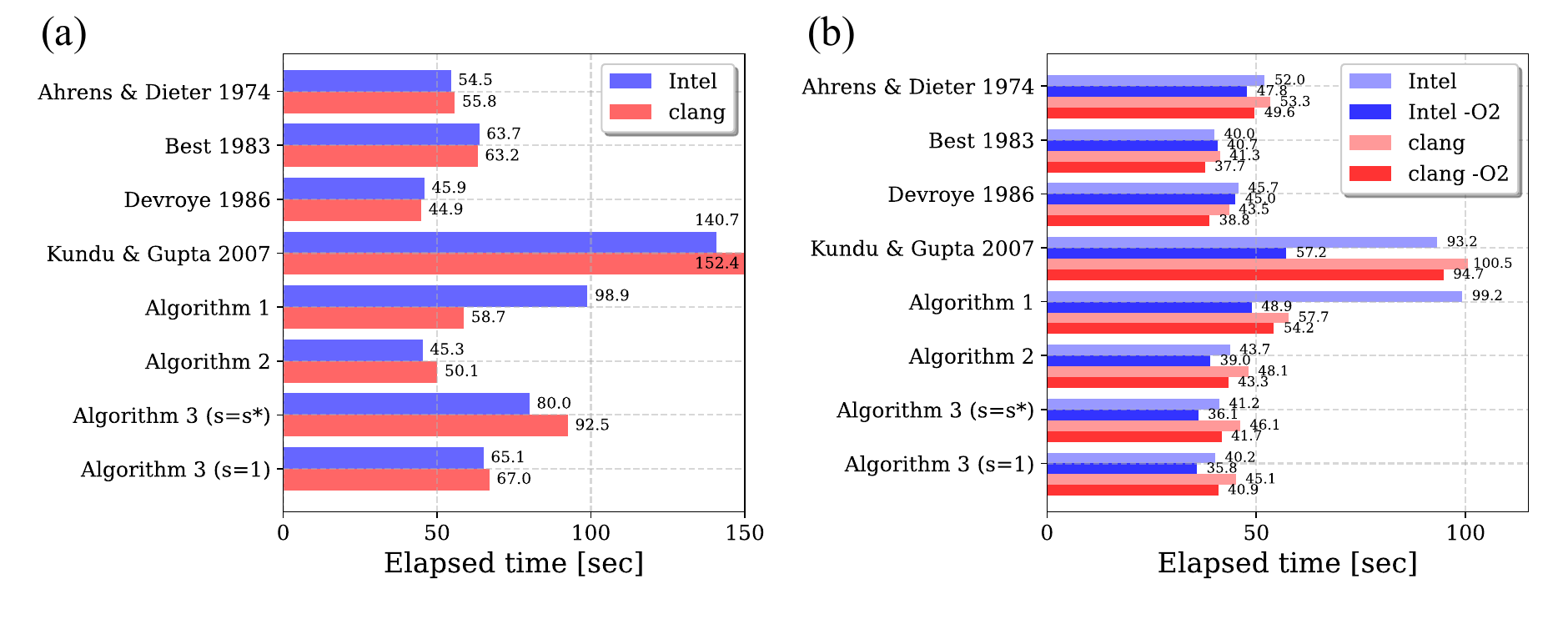}
\caption{
(a) Elapsed time to generate $10^8$ gamma variates, including the parameter initialization.
(b) Elapsed time to generate $10^8$ gamma variates all at once.
See the main text for detail.
\label{fig:bar}}
\end{figure*}

Next we compare the speed of the algorithms.
Considering practical applications,
two settings are considered \citep{tanizaki08}.
In the first setting, we repeatedly
draw one random variate $10^8$ times for all 11 parameters. 
We measure the elapsed time in seconds three times per method,
and then take the average elapsed time.
Figure \ref{fig:bar}(a) shows the results. 
The \citet{devroye86} method and Algorithm 2 are the fastest.
They outperform the well established methods \citep{AD74,best83}.
Although the two compilers give different results, 
from Algorithms 1 and 2, we see that the squeeze technique is effective.
Note that the \citet{kundu07} method does not use the squeeze technique.

In the second setting, we draw $10^8$ random variates all at once,
and then repeat this procedure for the 11 parameters. 
Some algorithms initialize internal parameters,
before generating many random variates.
For example, in Algorithm 3, the first three lines in Table \ref{table:GE3}
are used only once per parameter, before generating many variates at once.
These lines are repeated each time we draw a variate in the first setting.
Figure \ref{fig:bar}(b) shows the elapsed times.
For reference, results with high-level compiler optimizations
(the {\tt -O2} option) are also displayed. 
Now the piecewise algorithms
such as the \citet{best83} method and Algorithm 3 become faster
than in Fig.~\ref{fig:bar}(a), suggesting that
they are fast, once the parameters are initialized.
Together with the \citet{best83} method and the \citet{devroye86} method,
Algorithms 2--3 are close and promising.
In Algorithm 3, the $s=1$ cases are always faster than the $s=s^{*}$ cases.


\section{Discussion and Summary}
\label{sec:discussion}

We have developed the rejection algorithms
to generate a gamma variate with the shape parameter $0<\alpha<1$.
In line of \citet{kundu07}, we have employed the GE distribution \citep{gupta99}.
The GE distribution is useful, because it resembles the gamma distribution, and because its CDF has a simple form. 
\citet{kundu07} employed the GE distribution
with the scale parameter $2$, that is, $f_{\rm GE}(x; \alpha, 2)$,
while
we employ the scale parameter $1$, $f_{\rm GE}(x; \alpha, 1)$,
to obtain better acceptance efficiency.
Therefore, the parameters and the rejection function in their study
are different from ours.
Despite a different strategy, our acceptance efficiency is
as good as the best one \citep{devroye86}.

To improve the performance,
we have constructed the squeeze functions in \eqref{eq:sandwitch1}.
As demonstrated, this makes the code faster.
One may notice that
the left-hand side (the lower bound) can be negative for $x> 4/\beta$,
but logically there is no problem. 
In addition, since $F_{\rm GE}(4/\beta;\alpha,1) \ge F_{\rm GE}(4;\alpha,1) = (1-e^{-4})^{\alpha} \gtrsim 0.982$, more than 98\% of the random numbers are found for $x < 4/\beta$, where the lower bound condition is useful.

We have considered piecewise extensions of our algorithms,
with an exponential tail \citep{AD74,best83,kundu07}.
This makes the acceptance efficiency even better. 
The squeeze functions are presented in \eqref{eq:sandwitch2}.
The left inequality (the lower bound) is identical to
one in the \citet{best83} method,
while the right one (the upper bound) is a new addition.
This addition actually makes the \citet{best83} method few percents faster (not shown).
As for Algorithm 3,
despite the near-optimum density $S(\alpha,s^{*})<S(\alpha,1)$,
the $s=1$ case is always faster than the $s=s^{*}$ case,
because $s=1$ simplifies the algorithm.
Therefore $s=1$ is our choice.
In the second setting (Fig.~\ref{fig:bar}(b)),
Algorithm 3 with $s=1$ is always faster than Algorithm 2,
owing to its good acceptance rate.
It is promising when drawing many variates at once. 
In the first setting (Fig.~\ref{fig:bar}(a)),
in agreement with \citet{tanizaki08},
several methods become slower due to the parameter initialization.
This is also the case for Algorithm 3.
In contrast, Algorithm 2 remains fast in the first and second settings,
because of its simplicity.

Considering several aspects, we conclude that
Algorithm 2 is one of the best two generators, together with the \citet{devroye86} method.
These two often outperform the others,
including the well established methods \citep{AD74,best83}. 
With Intel compiler, Algorithm 2 is the best, and 
the \citet{devroye86} method is the best with clang.
The proposed methods are simple enough, and easy to implement,
as shown in Tables \ref{table:GE1}--\ref{table:GE3}.
In addition to the compiler-dependence,
there can be a CPU-dependence (and possibly a GPU-dependence). 
Considering various runtime environments,
we want to have as many good algorithms as possible, and
ours are certainly one of them.


In summary, we have developed gamma variate generators for $0 < \alpha < 1$,
using the GE distribution, the squeeze technique, and the piecewise envelope functions.
The proposed methods are excellent in acceptance efficiency.
The numerical tests suggest that Algorithm 2 is
one of the best two generators for $0 < \alpha < 1$.
Algorithm 3 with $s=1$ can be a good option when drawing many variates at once.
We hope that the proposed methods are useful in practical applications.

\section*{Data Availability}

Program codes in C language are available from the author upon request.

\end{document}